# Secure Performance Analysis and Optimization for FD-NOMA Vehicular Communications

Lai Wei[1], Yingyang Chen[2,*], Dongsheng Zheng[1], Bingli Jiao[1]

[1] Modern Communications Research Institute, Peking University, Beijing 100871, China
[2] College of Information Science and Technology, Jinan University, Guangzhou 510632, China
* The Corresponding Author, E-mail: chenyy@jnu.edu.cn

**Abstract:** Vehicle-to-vehicle (V2V) communication appeals to increasing research interest as a result of its applications to provide safety information as well as infotainment services. The increasing demand of transmit rates and various requirements of quality of services (QoS) in vehicular communication scenarios call for the integration of V2V communication systems and potential techniques in the future wireless communications, such as full duplex (FD) and non-orthogonal multiple access (NOMA) which enhance spectral efficiency and provide massive connectivity. However, the large amount of data transmission and user connectivity give rise to the concern of security issues and personal privacy. In order to analyze the security performance of V2V communications, we introduce a cooperative NOMA V2V system model with an FD relay. This paper focuses on the security performance of the FD-NOMA based V2V system on the physical layer perspective. We first derive several analytical results of the ergodic secrecy capacity. Then, we propose a secrecy sum rate optimization scheme utilizing the instantaneous channel state information (CSI), which is formulated as a non-convex optimization problem. Based on the differential structure of the non-convex constraints, the original problem is approximated and solved by a series of convex optimization problems. Simulation results validate the analytical results and the effectiveness of the secrecy sum rate optimization algorithm.

## I. INTRODUCTION

As the 5th generation (5G) wireless communication technology develops rapidly, various key techniques provide access to higher spectral efficiency and massive connectivity. Recent advances in interference cancellation accelerate the development of full duplex (FD) transceivers performing concurrent transmission and reception on the same channel and in the same time slot [1], which doubles spectral efficiency under the assumption of perfect self-interference cancellation. Other than FD radio technique, non-orthogonal multiple access (NOMA) paves the way for allowing multiple users to share time and frequency resources in the same spatial layer via power domain or code domain multiplexing [2]. By introducing controllable multiuser interference, NOMA provides access to higher user connectivity, pushing through the limitation

> This paper focuses on the security performance of the FD-NOMA based V2V system on the physical layer perspective.

of the maximum user number in traditional orthogonal multiple access (OMA) schemes. Moreover, researchers have focused on combining FD and NOMA in data transmissions. For instance, in [3], Ding et al. studied the performance of FD-NOMA, where uplink and downlink NOMA transmissions are carried out at the same time, demonstrating that FD-NOMA outperforms half duplex (HD)-NOMA and OMA with co-channel interference suppressed effectively.

Besides the advances in physical layer technology, the development in communication techniques has raised significant interest and progress in applying communications under complicated environments and scenarios, among which vehicular communication is one of the most compelling applications [4]. Vehicular communication enables automobiles to communicate with each other efficiently and safely through vehicle-to-vehicle (V2V) communications. Furthermore, vehicular communication also provides vehicle-to-infrastructure (V2I) communications, and generally, vehicle-to-everything (V2X) communications [5]. Up till now, there are two regimes in vehicular communication, namely the dedicated short-range communications (DSRC) [6], [7] and cellular-V2X (C-V2X) [8]. These days, C-V2X has received much attention with growing automobiles connecting to the vehicular communication network (VANET). The current version of C-V2X (i.e., LTE-V2X [9]) supports direct communications mode, in which vehicles can exchange information with each other directly. However, there still remain problems for LTE-V2X to satisfy diverse quality of services (QoS) and latency requirements [10].

Taking advantage of FD and NOMA, there exist some works on utilizing the aforementioned two techniques in vehicular communication scenarios for enhancing user connectivity and achieving higher data rates. In [11], Bazzi et al. investigated the use of FD radios for the beaconing transmission service among vehicles, demonstrating that the direct mode with FD saves the downlink channel resources and reduces the occupation of the uplink resources. Moreover, in [12], Di et al. studied the scheduling and resource allocation problem in a C-V2X broadcasting system and introduced a NOMA- based mixed centralized/distributed scheme to reduce the access collision and improve the network reliability. Besides, Zhang et al. [8] classified the V2X communications into the urban and crowd scenario as well as the suburban and remote scenario, introducing different channel models and deriving the exact system capacity expressions in both typical scenarios. In addition, Liu et al. [13] considered power allocation problems of HD relaying NOMA and FD relaying NOMA broadcasting/multicasting transmission schemes in 5G C-V2X communications to investigate and guarantee the QoS of vehicles with poor channel conditions.

Except for the benefits of FD-NOMA based vehicular communication hereinbefore, there exist concerns on the security issues of VANET. Therefore, it is a critical point for vehicular communication to address and account for security requirements prior to deployment [14]. Moreover, due to everything communicating to vehicles, a large amount of data is generated, which makes the privacy of users another big challenge in intelligent vehicle systems [15]. Current research mostly presents discussions on Internet oriented secure automotive technology and services, providing valuable perspectives and different dimensions to evaluate security issues of vehicular communication, including authentication, integrity, availability, confidentiality and physical security [16]. For instance, in [17], researchers discussed various security requirements of the users in VANET from the network perspective and considered various aspects of security attacks, entities and attackers. However, leveraging encryption techniques and de- signing security protocol to cope with security attacks may be not suitable for deployment in the vehicle environment since they do not support real-time data processing and are vulnerable to the attacks from external networks [18]. As has been discussed in [19], adopting generalized wireless authentication is an effective method to deal with security issues in smart vehicles. Nevertheless, there is also tremendous pressure to deploy certified identity for each vehicle as well as design the corresponding network protocol. All in all, the aforementioned works and recent research have informed us that it would be worthwhile considering the security issues of smart vehicles from the perspective of not only network or cryptography technology, but also the physical layer security (PLS).

It is well acknowledged that the concept of PLS was first discussed by Wyner [20]. The physical layer security schemes exploit the difference between legitimate and eavesdropping wireless links to improve communication security without using encryption technologies. There have been some works discussing PLS in NOMA systems. Researchers in [21] studied the secrecy sum rate maximization problem in a single-input- single-output (SISO) NOMA system with multiple legitimate users and one eavesdropper. Moreover, the PLS issue of cooperative NOMA systems for both amplify-and-forward (AF) and decode-and-forward (DF) protocols was analyzed in [22]. Besides, researchers have also investigated the PLS of applying NOMA in wireless networks [23], [24], [43]-[50]. Furthermore, there also exist several works on the performance of FD relay-assisted cooperative NOMA systems. Specifically, an FD relay assisted NOMA system was considered in [25], where the authors analyzed the outage probability and sum rate of the system over Nakagami-$m$ fading channel. Similarly, in [26], the authors proposed a dual user NOMA system with an FD relay assisting the information transmission to the user with weak channel condition and analyzed the achievable outage probability of both users and ergodic sum capacity.

However, to the best of the authors' knowledge, there are limited works considering PLS in vehicular communication, especially on the secrecy capacity analysis and optimization design. In [27], the authors analyzed the secrecy outage probability (SOP) of a cooperative NOMA vehicular communication system with either an HD or FD relay under Nakagami-$m$ fading channels. Besides, the authors of [28] investigated the joint relay selection and sub-carrier allocation problem for DF relay assisted V2V communications, minimizing the SOP of each user pair.

In this paper, we consider a cooperative NOMA V2V system with an FD relay and an eavesdropper. The ergodic secrecy capacity of such system is derived and several analytical expressions are presented. Furthermore, we propose a secrecy sum rate optimization problem utilizing the instantaneous channel state information (CSI). Numerical results also illustrate the effect of several system parameters on secrecy capacity and the effectiveness of the optimization scheme is also demonstrated. To sum up, the main contributions of this paper are three-fold:

- Firstly, we consider a basic FD-NOMA vehicular communication system consisting of a source vehicle, a relay vehicle working in FD mode, a NOMA vehicle user pair and an eavesdropper. The different statistical properties of each communication link are modeled, while both large- scale and small-scale fading are considered.
- Secondly, under the worst case assumption that the eaves- dropper has strong signal detection ability, we derive the analytical expressions of the achievable ergodic capacity of each NOMA user and the upper bound on the eavesdropping capacity. A lower bound on the ergodic secrecy capacity is then formulated. Simulation results demonstrate that the analytical derivation is approximate to the exact ergodic secrecy sum capacity. Moreover, the impact of several system parameters is evaluated in numerical analysis.
- Thirdly, considering the scenario that the CSI is obtained at source and relay vehicles, we propose a secrecy sum rate maximization problem with joint optimization of the power allocation parameters of source and relay. Our numerical results verify the effectiveness of the proposed optimization.

The rest of this paper is organized as follows. In Section II, the system model of FD-NOMA vehicular communication is presented. Several analytical results of ergodic secrecy capacity are derived in Section III, while in Section IV, we propose the secrecy sum rate optimization scheme. Besides, the numerical results are provided in Section V. Finally, the paper is concluded in Section VI.

*Notation*: Superscripts $[\cdot]^{ub}$ and $[\cdot]^{lb}$ denote an upper bound and lower bound on a function, respectively. $\mathbb{E}(\cdot)$ represents the mathematical expectation of a random variable. In this paper, we use $\gamma_i^j$ to denote the signal-to-interference-and-noise-ratio (SINR) for user $j$ to detect user $i$'s signal. Finally, $x \sim CN(\mu, \sigma^2)$ indicates that random variable $x$ obeys circularly symmetric complex Gaussian distribution with mean $\mu$ and variance $\sigma^2$.

## II. SYSTEM MODEL

As shown in Figure 1, we consider an FD-NOMA based V2V communication scenario consisting of five vehicles, one of which (source vehicle $S$) transmits message in the HD mode to a pair of users (user $D_1$ and user $D_2$) with the assistance of another vehicle (relay vehicle $R$) operating as a decode-and-forward (DF) FD relay, while an eavesdropper E is overhearing the message both from the source and the relay. As FD technique is employed, the relay vehicle suffers from self-interference (SI) that cannot be mitigated completely. Moreover, we assume that there are no direct links between source and destinations. Besides, in this paper, we assume that channels between any two vehicles experience block Rayleigh fading such that the channel coefficients vary independently from one block to another but remain constant over one transmission block. Such a channel model is also exploited in [8], moreover, the validation of Rayleigh fading is due to the abundant scatters between source and destination especially in urban and crowed area according to [29]. Specifically, the channel coefficients of source-relay $(S-R)$ link, relay-strong user $D_1$ $(R-D_1)$ link, relay-weak user $D_2$ $(R-D_2)$ link, source-eavesdropper link $(S-E$ link) and relay-eavesdropper $(R-E$ link) are denoted as $h_{S,R} \sim CN(0, d_{S,R}^{-v})$, $h_{R,D_1} \sim CN(0, d_{R,D_1}^{-v})$, $h_{R,D_2} \sim CN(0, d_{R,D_2}^{-v})$, $h_{S,E} \sim CN(0, d_{S,E}^{-v})$ and $h_{R,E} \sim CN(0, d_{R,E}^{-v})$, respectively, where $d_{i,j}$ represents the distance in meters between vehicle $i$ and vehicle $j$, and $v$ denotes the pathloss exponential parameter. For convenience, we also define that $d_{i,j}^{-v} = \sigma_{i,j}^2$. We then denote the additive white Gaussian noises (AWGNs) at $S, R, D_1, D_2$ and $E$ as $n_S, n_R, n_{D_1}, n_{D_2}$ and $n_E$, respectively. Without loss of generality, we assume that all AWGNs are independent and identically distributed (i.i.d) with a distribution of $CN(0, \sigma_n^2)$. Finally, the channel coefficient of the relay self-interference channel is denoted as $h_{R,R} \sim CN(0, \sigma_{SI}^2)$.

Suppose DF protocol is applied at the FD relay. Specifically, the FD relay decodes its received signal under the NOMA protocol and then re-encodes and forwards the signal of the user pair. Therefore, the instantaneous SINRs at the FD relay to decode symbols of the two users are respectively given by

$$\gamma_{D_2}^R = \frac{(1-a_S)\rho |h_{S,R}|^2}{a_S\rho |h_{S,R}|^2 + \rho_{SI}|h_{R,R}|^2 + 1}, \quad (1)$$

$$\gamma_{D_1}^R = \frac{a_S\rho |h_{S,R}|^2}{\rho_{SI}|h_{R,R}|^2 + 1}. \quad (2)$$

where $\rho = \frac{P}{\sigma_n^2}$ and $\rho_{SI} = \frac{P_{SI}}{\sigma_n^2} = \frac{k_r P}{\sigma_n^2}$, with P denoting the transmit power and $k_r$ defining the level of residual SI under imperfect interference cancellation.- In this article, we use $a_S$ and $a_R$ denoting the power allocation parameter at source and relay vehicle, according to NOMA protocol, we set $0 < a_S, a_R < \frac{1}{2}$ by default. Moreover, according to the NOMA protocol, after receiving signals from the FD relay, user $D_1$ firstly decodes the symbol of user $D_2$ before decoding that of itself, while user $D_2$ decodes its own symbol directly. Therefore, the SINR at the two users to decode each symbol, namely $\gamma_{D_1}^{D_1}, \gamma_{D_2}^{D_1}, \gamma_{D_2}^{D_2}$ can be derived in the similar procedure.

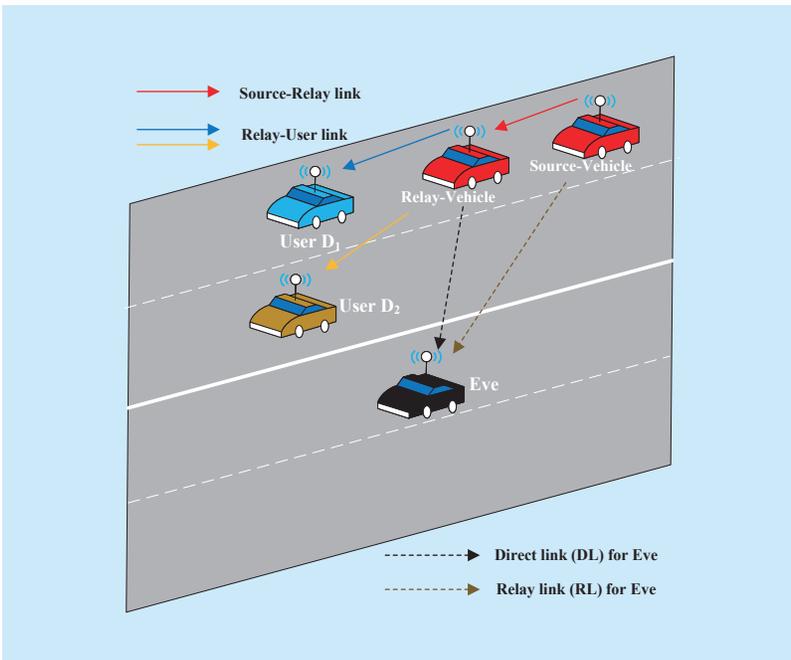

**Fig. 1.** *FD-NOMA vehicular communication system.*

According to [30] and [31], the $R-E$ link from the relay to the eavesdropper vehicle has insignificant time delay compared with the $S-E$ link from source to Eve. In other words, the signal from the $R-E$ link and that of $S-E$ link can be resolvable simultaneously for the eavesdropper. Suppose that the eavesdropper decodes the receiver superimposed signal using the NOMA protocol as well. Throughout this paper, we make a pessimistic assumption that the eavesdropper has strong signal detection ability. Specifically, we assume that the eavesdropper is able to collect the signals from $S-E$ link and $R-E$ link using maximum ratio combining (MRC) [32] and decode the superimposed signal by exploiting NOMA. Thus, the SINRs at the eavesdropper to detect the users' messages after MRC can be further expressed as [33]

$$\gamma_{D_2}^E = \frac{(1-a_S)\rho |h_{S,E}|^2 + (1-a_R)\rho |h_{R,E}|^2}{a_S\rho |h_{S,E}|^2 + a_R\rho |h_{R,E}|^2 + 1}, \quad (3)$$

$$\gamma_{D_1}^E = a_S\rho |h_{S,E}|^2 + a_R\rho |h_{R,E}|^2 + 1, \quad (4)$$

respectively. Finally, we define the effective SINR of the two users as $\gamma_{D_1} = \min\left(\gamma_{D_1}^R, \gamma_{D_1}^{D_1}\right)$ and $\gamma_{D_2} = \min\left(\gamma_{D_2}^R, \gamma_{D_2}^{D_1}, \gamma_{D_2}^{D_2}\right)$.

## III. ERGODIC SECRECY SUM RATE ANALYSIS

In this section, we analyze the ergodic secrecy sum rate of the proposed FD-NOMA based V2V system. According to [34], the ergodic secrecy capacity of user $i$ is given by

$$\overline{C}_i^{sec} = [\overline{C}_i - \overline{C}_i^E]^+, \quad (5)$$

where $\overline{C}_i$ and $\overline{C}_i^E$ are the ergodic achievable rate and the ergodic eavesdropping rate of the $i$-th user respectively, and $[\cdot]^+$ equals $\max(\cdot, 0)$.

To obtain the ergodic secrecy capacity formula, we first investigate the distribution function of the received SINR of each symbol at each vehicle. The main results are as follows.

*Theorem* 1: The ergodic achievable capacities of the two users are given by

$$\overline{C}_{D_1} = \frac{c_1}{(c_1-c_2)\ln 2}[e^{\frac{1}{c_1}}E_1(\frac{1}{c_1}) - e^{\frac{1}{c_2}}E_2(\frac{1}{c_2})], \quad (6)$$

$$\overline{C}_{D_2} = \frac{1}{\ln 2}\int_0^{+\infty} \frac{1}{1+x} \cdot \frac{1}{1 + \frac{\lambda_{S,R}\rho_{SI}x}{\lambda_{R,R}(1-a_S-a_S x)\rho}} \quad (7)$$

$$\cdot e^{-\frac{\lambda_{S,R}x}{(1-a_S-a_S x)\rho}} \cdot e^{-\frac{(\lambda_{R,D_1}+\lambda_{R,D_2})x}{(1-a_R-a_R x)\rho}} dx.$$

respectively, where the constants $c_1 = \frac{1}{\pi_{S,R}+\pi_{R,D_1}}$, $c_2 = \frac{\rho_{SI}\pi_{S,R}}{\lambda_{R,R}\pi_{S,R}+\pi_{R,D_1}}$ and $\lambda_{i,j} = \frac{1}{\sigma_{i,j}^2}, \pi_{S,R} = \frac{1}{a_S\rho\sigma_{S,R}^2}, \pi_{R,D_1} = \frac{1}{a_R\rho\sigma_{R,D_1}^2}$.

The function $E_1(z) = -\text{Ei}(-z) = \int_z^{+\infty} \frac{e^{-t}}{t}dt$, with $\text{Ei}(z)$ denoting the exponential integral [35].

*Proof*: By definition of ergodic capacity and with the help of integral by parts, the ergodic capacity of strong user $D_1$ can be expressed as

$$\overline{C}_{D_1} = \mathbb{E}[\log(1+\gamma_{D_1})] = \frac{1}{\ln 2}\int_0^{+\infty} \ln(1+x)f_{\gamma_{D_1}}(x)dx$$

$$= \frac{1}{\ln 2}\int_0^{+\infty} \frac{1-F_{\gamma_{D_1}}(x)}{1+x}dx, \quad (8)$$

where the cumulative distribution function

(c.d.f.) of $\gamma_{D_1}$ could be calculated as

$$F_{\gamma_{D_1}}(x) = 1 - P(\gamma_{D_1} \geq x)$$
$$= 1 - P(\gamma_{D_1}^R \geq x) \cdot P(\gamma_{D_1}^{D_1} \geq x). \#$$

Therefore, the c.d.f. of SINRs can be characterized by

$$F_{\gamma_{D_1}}(x) = 1 - \frac{\lambda_{R,R} e^{-(\pi_{S,R} + \pi_{R,D_1})x}}{\rho_{SI}\pi_{S,R}x + \lambda_{R,R}},$$

$$F_{\gamma_{D_2}}(x) = 1 - e^{-\frac{\lambda_{S,R}x}{(1-a_S-a_Sx)\rho}} \cdot e^{-\frac{(\lambda_{R,D_1}+\lambda_{R,D_2})x}{(1-a_R-a_Rx)\rho}}$$

$$\cdot \frac{1}{1 + \frac{\lambda_{S,R}\rho_{SI}x}{\lambda_{R,R}(1-a_S-a_Sx)\rho}},$$

respectively.

Substituting the c.d.f. of $\gamma_{D_1}$ into (8), and after some algebraic manipulations, we obtain the desired results. The ergodic achievable capacity of user $D_2$ can be yielded by using the same calculation procedure.

Similar to the analysis of the ergodic achievable capacities of the two users, the ergodic eavesdropping capacity of the strong user $D_1$ is given in the following theorem.

*Theorem* 2: The ergodic eavesdropping capacity of the strong user $D_1$ is given by

$$\bar{C}_{D_1}^E = \frac{(\pi_{R,E} e^{\pi_{S,E}} E_1(\pi_{S,E}) - \pi_{S,E} e^{\pi_{R,E}} E_1(\pi_{R,E}))}{(\pi_{R,E} - \pi_{S,E})\ln 2}, (9)$$

with $\pi_{S,E} = \frac{1}{a_S\rho\sigma_{S,E}^2}$ and $\pi_{R,E} = \frac{1}{a_R\rho\sigma_{R,E}^2}$. Additionally, considering the case that the direct link and the eavesdropping link have the same channel condition, the ergodic eavesdropping capacity could be derived by calculating the mathematical expectation on a Gamma distribution function [32].

*Proof*: The p.d.f.s of eavesdropping SINR of user $D_1$ can be characterized as

$$f_{\gamma_{D_1}^E}(x) = \frac{\pi_{S,E}\pi_{R,E}}{\pi_{R,E} - \pi_{S,E}}(e^{-\pi_{S,E}x} - e^{-\pi_{R,E}x})$$

Substituting the p.d.f. into (9), we derive the desired results.

On the other hand, the exact mathematic expression of the ergodic eavesdropping capacity of the weak user $D_2$ is hard to obtain due to the asymmetric structure of the numerator and denominator of the instantaneous SINR.

Therefore, we propose the following upper bound to approximate the ergodic eavesdropping capacity of user $D_2$.

*Theorem* 3: An upper bound on the ergodic eavesdropping capacity of user $D_2$ takes the following form

$$\bar{C}_{D_2}^{E,ub} = \frac{1}{\ln 2} \int_0^{+\infty} \frac{\pi_{R,E} e^{-\frac{\pi_{S,E}x}{(1-a-ax)\rho}} - \pi_{S,E} e^{-\frac{\pi_{R,E}x}{(1-a-ax)\rho}}}{(1+x)(\pi_{R,E} - \pi_{S,E})} dx,$$
(10)

where $a = \min(a_S, a_R)$.

*Proof*: Note that
$$\gamma_{D_2}^E \leq \delta_{D_2}^E = \frac{(1-a)\rho(|h_{S,E}|^2 + |h_{R,E}|^2)}{a\rho(|h_{S,E}|^2 + |h_{R,E}|^2) + 1}.$$

Therefore, an upper bound on user $D_2$'s eavesdropping capacity can be obtained as $\bar{C}_{D_2}^E \leq \mathbb{E}\left[\log(1+\delta_{D_2}^E)\right] = \bar{C}_{D_2}^{E,ub}$. Specifically, the p.d.f. of $\delta_{D_2}$ and its mathematical expectation can be calculated in the similar procedure to that performed in *Theorem* 2.

To the best of our knowledge, it is difficult to formulate the integrals hereinbefore into close-form solutions. However, the integrals can be calculated by numerical integration, which is more efficient than the time-consuming Monte Carlo method [26]. Therefore, the ergodic secrecy sum rate can be derived by calculating and substituting the aforementioned integrals into its definition expression. To sum up, we conclude the results in this section into the following theorem.

*Theorem* 4: A lower bound on the ergodic secrecy sum rate of the FD-NOMA V2V communication system can be expressed as

$$\bar{C}_{sec} \geq \bar{C}_{sec}^{lb} = [\bar{C}_{D_1} - \bar{C}_{D_1}^E]^+ + [\bar{C}_{D_2} - \bar{C}_{D_2}^{E,ub}]^+. (11)$$

## IV. SECRECY SUM RATE MAXIMIZATION

In this section, we consider that S and R know the instantaneous CSI of the legitimate link as well as that of the eavesdropping links. Therefore, the source and relay vehicle will be able to jointly optimize the power allocation parameter to maximize the secrecy sum rate. Similar to the aforementioned analysis, the secrecy sum rate maximization problem can be denoted as the following non-convex optimization problem

$$\max_{a_S, a_R} R_{\text{sec}} = \sum_{i \in \{D_1, D_2\}} \max(0, R_i - R_i^E)$$
$$s.t. \ 0 \leq a_S \leq \frac{1}{2}, 0 \leq a_R \leq \frac{1}{2}. \quad (12)$$

Note that the instantaneous secrecy rate can be reformulated into a difference-of-two-concave function (DC) form. Specifically, for each user we obtain

$$R_{D_1} - R_{D_1}^E = \min(\log\frac{1+\gamma_{D_1}^R}{1+\gamma_{D_1}^E}, \log\frac{1+\gamma_{D_1}^{D_1}}{1+\gamma_{D_1}^E})$$
$$= \min(\psi_1 - \phi_1, \pi_1 - \omega_1),$$
$$R_{D_2} - R_{D_2}^E = \min(\log\frac{1+\gamma_{D_2}^R}{1+\gamma_{D_2}^E}, \log\frac{1+\gamma_{D_2}^{D_1}}{1+\gamma_{D_2}^E}, \log\frac{1+\gamma_{D_2}^{D_2}}{1+\gamma_{D_2}^E})$$
$$= \min(\psi_2 - \phi_2, \pi_2 - \omega_2, \theta_2 - \xi_2). \quad (13)$$

Therefore, by introducing several auxiliary variables, the optimization problem can be reformulated into an equivalent form

$$\max_{a_S, a_R, t_1, t_2} t_1 + t_2$$
$$s.t. \ 0 \leq a_S \leq \frac{1}{2}, 0 \leq a_R \leq \frac{1}{2},$$
$$t_1 \geq 0, t_2 \geq 0, \quad (14)$$
$$t_1 \leq \psi_1 - \phi_1, t_1 \leq \pi_1 - \omega_1,$$
$$t_2 \leq \psi_2 - \phi_2, t_2 \leq \pi_2 - \omega_2, t_2 \leq \theta_2 - \xi_2.$$

Moreover, with the help of the DC structure of the objective function hereinbefore, the non-convex optimization problem can be transformed into a convex optimization problem. Specifically, as shown in [36], we use the first order approximation of the constraints, namely, $f(x) \approx f(x_0) + \nabla f(x_0)^T (x - x_0)$ to and solve the optimization problem in each step of iteration, so that the primal constraints of problem become convex functions.

Therefore, the problem can be solved by successively handling a series of convex optimization problems. As the iteration proceeding, an optimal value of the objective function can be obtained and the convex approximation generated by DC programming can successively converge to the original function [37], [38]. Herein, we use $a_S^{(l)}$ and $a_R^{(l)}$ to denote the power allocation parameters after $l$ iterations. Finally, we clarify the DC-programming based secrecy sum rate optimization problem in **Algorithm 1**.

---

**Algorithm 1.** DC-Programming Based Secrecy Sum Rate (SSR) Optimization Problem.

**Input:** Instantaneous CSI and error tolerance $\epsilon \ll 1$,
**Output:** Optimal $a_S^{(opt)}$, $a_R^{(opt)}$ and maximum SSR

1. Iterator $l = 1$;
2. Randomly choose $a_S^{(l-1)}$ and $a_R^{(l-1)}$ in the feasible set;
3. Set $SSR^{(l-1)} = 0$;
4. **Repeat:**
5. Calculate values of the function $\phi_1, \omega_1, \phi_2, \omega_2, \psi_2$ and their first order derivatives at $\left(a_S^{(l-1)}, a_R^{(l-1)}\right)$;
6. Substituting $\phi_1, \omega_1, \phi_2, \omega_2$ and $\theta_2$ with their first order approximations;
7. Solve the convex optimization problem and obtain $a_S^{(l)}, a_R^{(l)}$ and $t_1, t_2$;
8. Set $a_S^{(l-1)} = a_S^{(l)}$ and $a_R^{(l-1)} = a_R^{(l)}$;
9. Set $SSR^{(l)} = \max\left(SSR^{(l-1)}, t_1 + t_2\right)$;
10. Set $l = l + 1$;
11. **Until:** $\left|a_S^{(l)} - a_S^{(l-1)}\right| < \epsilon, \left|a_R^{(l)} - a_R^{(l-1)}\right| < \epsilon$;
12. Set $a_S^{(opt)} = a_S^{(l)}$ and $a_R^{(opt)} = a_R^{(l)}$;
13. Maximum $SSR = SSR^{(l)}$.

---

## V. SIMULATION RESULTS

In this section, we provide simulation results to evaluate and demonstrate the secrecy performance of the FD-NOMA based V2V system. Firstly, we analyze the ergodic secrecy capacity of the system under several different circumstances. Then, we present the simulation results of the proposed secrecy sum rate optimization scheme.

### 5.1 Ergodic secrecy sum rate analysis

For analyzing ergodic secrecy sum rate, the system parameter configurations in different scenarios are clarified as follows. The pathloss exponential parameter is set as $v = 3$. Moreover, we assume that the distance between source and relay vehicles is 10 m, while the relay vehicle is 10 m and 15 m apart from the strong user $D_1$ and the weak user $D_2$, respec-tively, which is very typical for urban area especially in rush hours [39].

In Figs. 2 and 3, we focus on the effect of the power allocation parameters on the secrecy sum rate. Under such circumstances, the normalized transmit signal-to-noise-ratio (SNR) is set as $\rho=30\,dB$, and the residual SI power is set as $\rho_{SI}=-10\,dB$. Meanwhile, we assume that the eavesdropper is relatively far. Specifically, the distance between Eve and the source vehicle is set as $d_{S,E}$ = 40 m, while the relay vehicle is 30 meters away from Eve, namely $d_{R,E}$ = 30 m. In the first stage, we fix the power allocation parameter of the relay vehicle as $a_R=0.14$. Then the relationship between $a_S$ and ergodic secrecy sum rate is presented. Similarly, the effect of $a_R$ on the ergodic secrecy capacity is evaluated vice versa. Simulation results are obtained using classical Monte Carlo method with $10^6$ channel realizations.

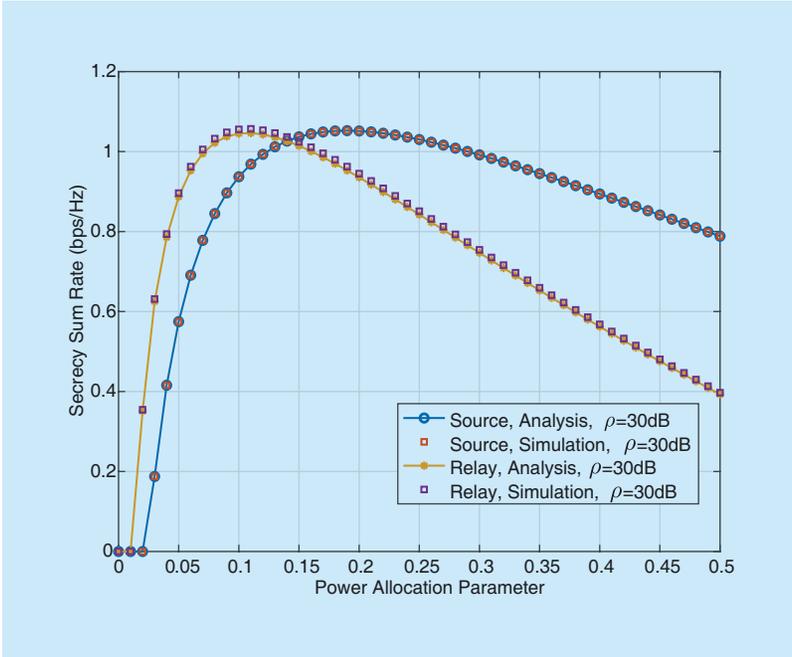

**Fig. 2.** *Ergodic Secrecy Sum Rate versus power allocation parameter (high SNR regime, $\rho = 30dB$).*

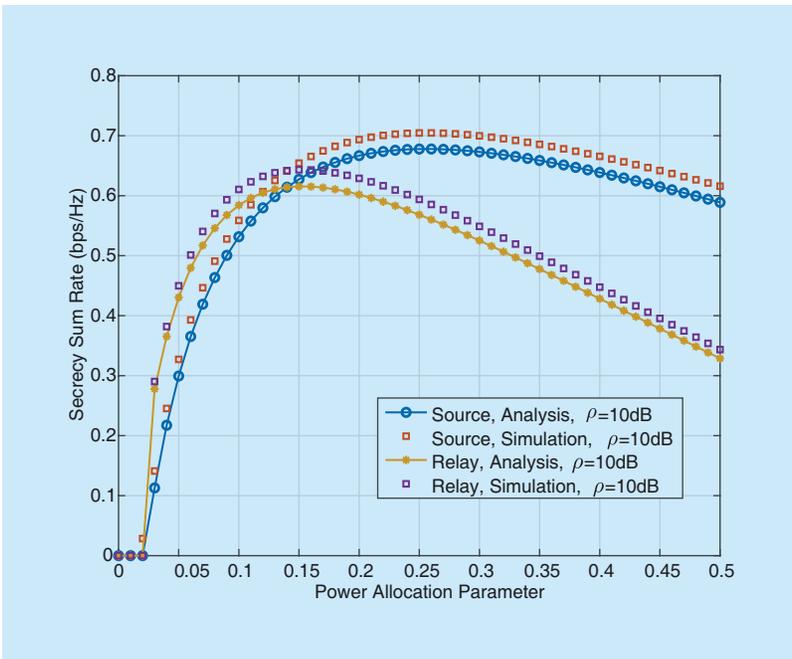

**Fig. 3.** *Ergodic Secrecy Sum Rate versus power allocation parameter (lower SNR regime, $\rho =10dB$).*

Figs. 2 and 3 show that the ergodic secrecy sum rate function has a concave structure with respect to power allocation parameter $a_S$ and $a_R$, which implies that one can design a gradient ascent algorithm for source and relay vehicles to op- timize ergodic secrecy sum rate. However, due to the complex mathematical expression of each user's ergodic secrecy capac- ity, it is difficult to obtain the first or second order derivative of the ergodic secrecy capacity function with respect to $a_S$ or $a_R$. Moreover, it is also difficult to determine the concave structure of ergodic secrecy capacity function analytically. It is worth pointing out that, similar to the results obtained in [40], whether the ergodic secrecy capacity is always concave with respect to $a_S$ and $a_R$ is an open problem. Besides, Figure 2 implies that the lower bound obtained in Section III is tight in the high SNR regime. Moreover, Figure 3 illustrates that as the power allocation parameters increase, the analytical bound also shows a similar trend to that of the simulation results in the lower SNR regime, and the gap between analytical and simulation results is also relatively small.

The ergodic secrecy capacity versus the distance of the eavesdropper is depicted in Figure 4. The residual SI power and the power allocation parameters are set as $\rho_{SI}=-10\,dB$ and $a_S=a_R=0.20$ by default, and we assume that $d_{R,E}=\frac{1}{2}d_{S,E}$. Notice that the ergodic se-crecy capacity has a trend of increasing first and then becomes steady as the eavesdropper moves away. The reason is that the influence of eavesdropper can be neglected as Eve moves far away, and thus the secrecy sum rate converges to the ergodic sum capacity of a two-user FD-NOMA V2V system.

Figure 5 illustrates the ergodic secrecy sum rate versus transmit power. The residual SI power is set as $\rho_{SI} = -10\,dB$, while $a_S = a_R = 0.20$ is fixed. Firstly, we analyze the system performance when the eavesdropper is close to the source and relay vehicles. To be specific, the distances between the eavesdropper and the source and relay vehicles are set as $d_{S,E} = 25\,m$ and $d_{R,E} = 20\,m$, respectively. Secondly, an FD-NOMA V2V system associated with a mid-range Eve with $d_{S,E} = 30\,m$ and $d_{R,E} = 20\,m$ are investigated. Finally, the eavesdropper is set to be further with $d_{S,E} = 40\,m$ and $d_{R,E} = 30\,m$.

It can be observed from Figure 5 that with a close eavesdropper, as the transmit power increases, secrecy sum rate declines on the contrary. As transmit power becomes high enough, the secrecy capacity even goes to zero, which makes it impossible for secure communications. Since the eavesdropper vehicle overhears and decodes the users' signal through MRC, the increasing transmit power is more favorable to Eve than the user vehicles if Eve is close to source and relay. As the eavesdropper moves further, the benefit of enlarging transmit power becomes dominant resulting in the increase of the secrecy sum rate.

In Figure 6, the secrecy ergodic rate associated with residual SI power is presented with $d_{S,E} = 40\,m$ and $d_{R,E} = 30\,m$. Clearly, the residual SI signal has significant negative effect on the system performance. Furthermore, we can infer from the blue plot that, the FD-NOMA vehicular system is able to obtain fair security performance if the SI capability is no less than 20 dB (that is, $\rho - \rho_{SI} \geq 20$ dB). According to [41]–[43], it is achievable for the FD radio system to suppress the SI up to 70-110 dB. Therefore, it is feasible for the FD-NOMA vehicular communication system to achieve secure communication. Additionally, with the help of SI cancellation technique, the proposed FD-NOMA scheme could outperform

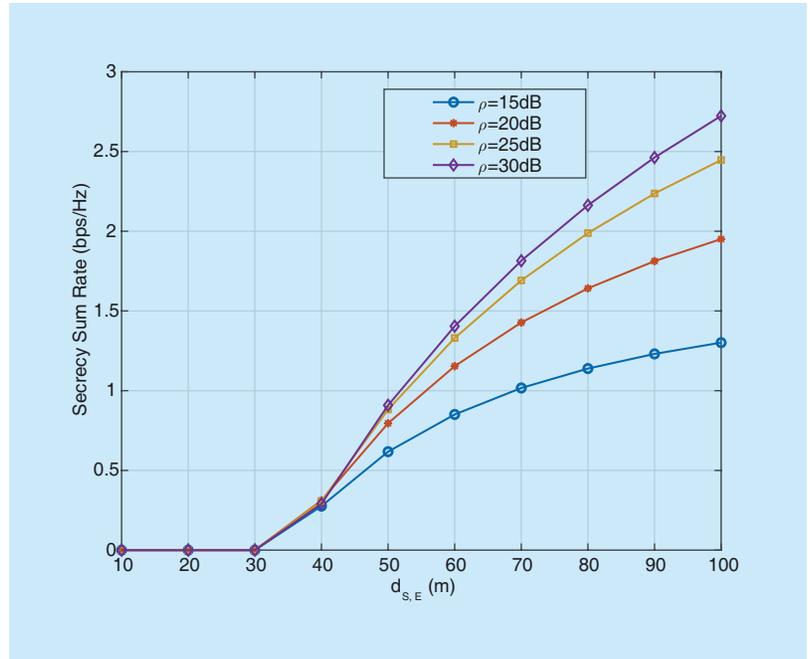

**Fig. 4.** *Ergodic secrecy sum rate performance with Eve at different distances.*

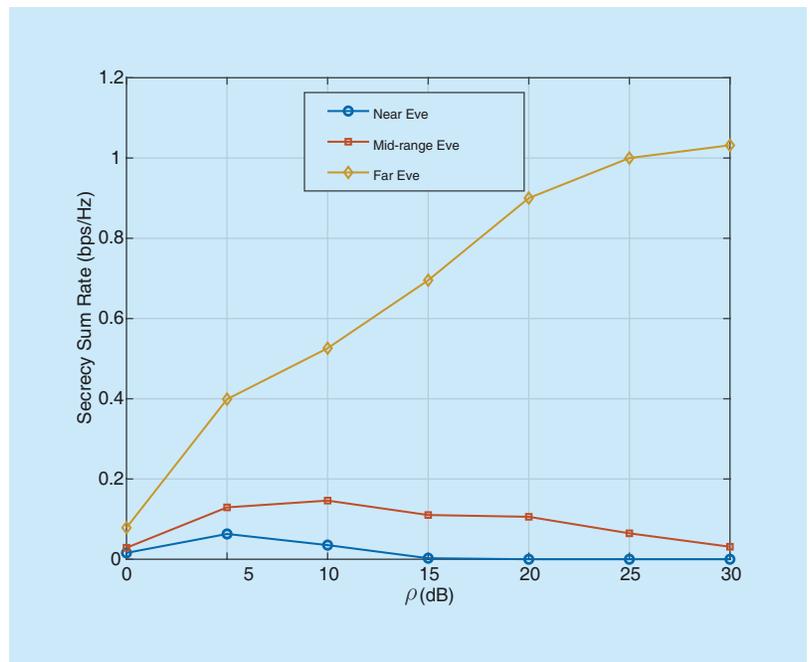

**Fig. 5.** *Ergodic secrecy sum rate versus transmit power.*

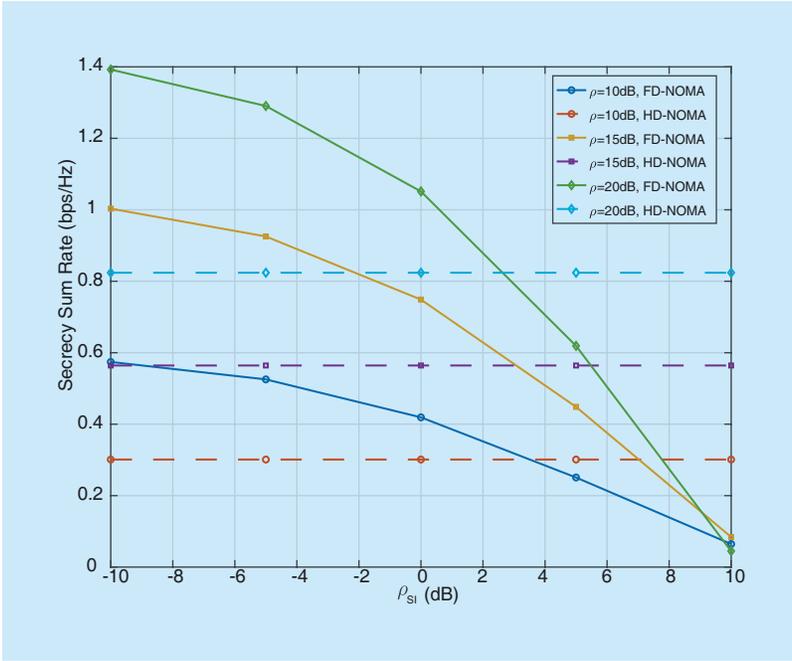

**Fig. 6.** *Ergodic secrecy sum rate versus residual SI power.*

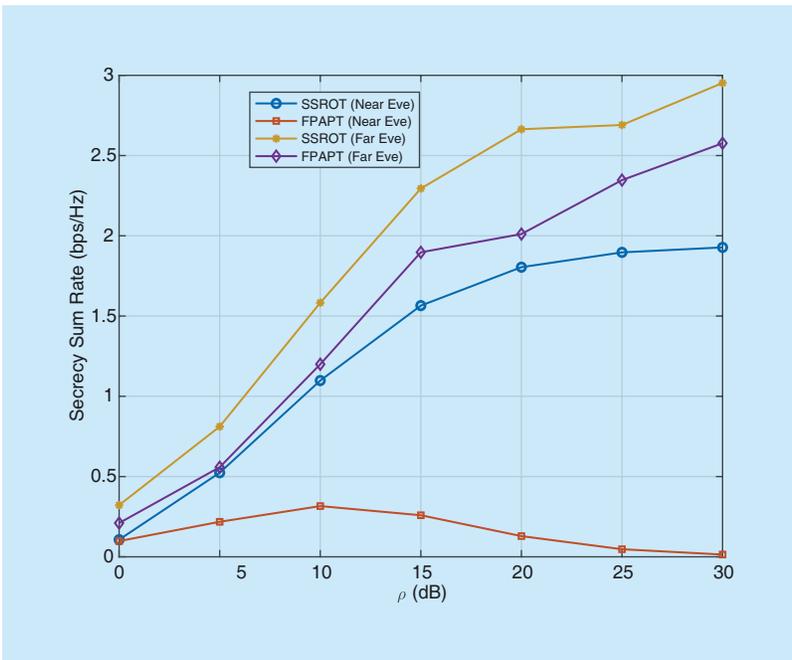

**Fig. 7.** *Performance analysis of the proposed secrecy sum rate optimization scheme.*

HD-NOMA scheme and the secrecy sum rate could be doubled if the SI signal is sufficiently suppressed. On the other hand, the equipment of the FD radio is not guaranteed to enhance secrecy capacity if the SI signal is not ideally cancelled.

## 5.2 Secrecy sum rate optimization scheme

In this subsection, we demonstrate the effectiveness of the proposed secrecy sum rate optimization scheme. The parameter configurations are summarized as follows. The distance parameters are set as $d_{S,E} = 40\,m$ and $d_{R,E} = 30\,m$ for a far eavesdropper, while $d_{S,E} = 25\,m$ and $d_{R,E} = 20\,m$ for a near Eve. We fix residual SI power as $\rho_{SI} = -10\,dB$. For the fixed power allocation parameter transmission (FPAPT) scheme, the power allocation parameters are fixed as $a_S = a_R = 0.20$. In addition, the instantaneous CSI is randomly generated per transmission and simulation results are obtained by using Monte Carlo method with $10^4$ channel realizations.

As shown in Figure 7, under the premise that the eavesdropper is relatively far, with the help of the proposed secrecy sum rate optimization transmission (SSROT) scheme, the FD-NOMA V2V system adaptively allocates the transmission power of the NOMA user pair, leading to a higher secrecy sum rate in each transmission than the FPAPT scheme. On the other hand, as the eavesdropper comes closer to the user, the proposed SSROT scheme effectively guarantees secure communications, while fixing power allocation parameter may lead to severe performance degradation.

## VI. CONCLUSION

This paper has investigated the ergodic secrecy sum rate of an FD-NOMA based V2V system. Analytical results on ergodic capacities of both NOMA user pair and eavesdropper have been derived. The impact of several factors including distance between Eve and the user pair, transmission SNR and residual SI power have been taken into account. Taking advantage of the DC structure of instantaneous secrecy sum rate, a secrecy sum rate optimization scheme with instantaneous CSI has been proposed.

Simulation results have demonstrated the effectiveness of the proposed scheme compared with fixed- power transmission.

请按我刊参考文献著录格式要求修改：刊名为斜体，刊出年月放在页码之前等。

## Biographies

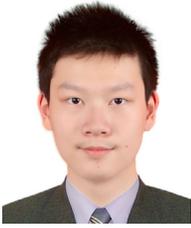
***Lai Wei,*** received the B.S. degree in Electronic Engineering from Peking University in 2018. He's currently pursuing the Ph.D. degree in signal and information processing with the Modern Communications Research Institute, Peking University, Beijing, China. His research interest includes wireless communication, signal processing and information theory. (E-mail: future1997@pku.edu.cn).

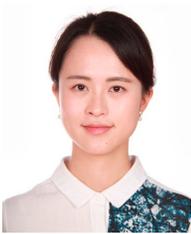
***Yingyang Chen [S'16],*** received the B.Eng. degree in electronic engineering from Yingcai Honors College, University of Electronic Science and Technology of China (UESTC), Chengdu, China, in 2014. She received her Ph.D. degree in signal and information processing from Peking University, Beijing, China, in 2019. From March 2018 to September 2018, she worked as a visiting student in the Next Generation Wireless Group at the University of Southampton, supervised by Prof. Lajos Hanzo. Her current research interests include performance analysis and optimization in wireless communication systems, mobile edge caching and computing, and array signal processing. (E-mail: chenyy@jnu.edu.cn).

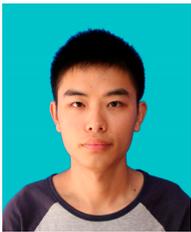
***Dongsheng Zheng,*** received the B.Eng. degree in microelectronic science and engineering from the University of Electronic Science and Technology of China, Chengdu, China, in 2018. He is currently pursuing the Ph.D. degree in signal and information processing with the Modern Communications Research Institute, Peking University, Beijing, China. His research interests include mobile edge caching, vehicular communications, and signal processing in wireless communications. (E-mail: zhengds@pku.edu.cn).

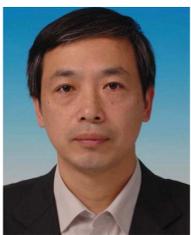
***BingLi Jiao [M'05, SM'11],*** received the B.S. and M.S. degrees from Peking University, China, in 1983 and 1988, respectively, and the Ph.D. degree from Saarland University, Germany, in 1995. Then, he became an associate professor in 1995 and a professor with Peking University in 2000. His current interests include full duplex communications, information theory, and signal processing. He is a director of the Joint Laboratory for Advanced Communication research between Peking University and Princeton University. He is a pioneer of co-frequency and co-time full duplex as found in his early patent in 2016. (E-mail: jiaobl@pku.edu.cn).